\documentclass[a4paper,10pt]{article}
\usepackage{amsmath,amssymb,epsfig,graphicx} 
\usepackage{setspace}

\begin{document}


\title{Precise simulation of the initial-state QCD activity 
associated with $Z$-boson production in hadron collisions}

\author{Shigeru Odaka\\
High Energy Accelerator Research Organization (KEK)\\
1-1 Oho, Tsukuba, Ibaraki 305-0801, Japan\\
E-mail: \texttt{shigeru.odaka@kek.jp}}

\date{}

\maketitle

\begin{abstract}
The $\phi_{\eta}^{*}$ distribution of the 
$Z/\gamma^{*} \rightarrow \ell^{+}\ell^{-}$ production in hadron collisions 
is simulated using a leading-order event generator, GR@PPA.
The initial-state parton shower, 
which simulates the multiple QCD-radiation effects in the initial state, 
plays the dominant role in this simulation.
The simulation in the default setting agrees 
with the high-statistics measurement by ATLAS at LHC with the precision 
at the level of 5\%.
The observed systematic deviation, 
which can be attributed to the effects of ignored higher-order contributions, 
can be reduced by adjusting the arbitrary energy scales in the simulation.
The agreement at the level of 1\% can be achieved over a very wide range 
without introducing any modification in the implemented naive 
leading-logarithmic parton shower.
\end{abstract}

\section{Introduction}
\label{sec:intro}

$Z$ boson production in hadron collisions is a unique place for 
studying the initial-state strong interaction (QCD) activity,
because the production kinematics can in principle be unambiguously measured 
if the production is tagged by a lepton pair (electrons or muons) 
from the decay.
The spectrum of the $Z$-boson transverse momentum, $p_{T}(Z)$,
with respect to the beam direction, 
which can be reconstructed from the measured decay-lepton momenta, 
is the most important property for probing the underlying QCD interactions.
We expect that the $p_{T}(Z)$ spectrum at high $p_{T}(Z)$ ($\gtrsim m_{Z}$) 
can be well described by the production of a high-$p_{T}$ jet 
(light quark or gluon) in association with the $Z$ production, 
where $m_{Z}$ represents the invariant mass of the $Z$ boson.
However, as $p_{T}(Z)$ decreases, the contribution of the multiple QCD 
radiation becomes significant and alters the naive prediction from the 
$Z$ + 1 jet production.
Finally, at $p_{T}(Z) < 10 {\rm ~GeV}/c$, the divergent $Z$ + 1 jet 
cross section is suppressed to form a peak structure at several GeV/$c$.
A certain non-perturbative contribution is considered to be important 
around the peak.
These features must basically be common to all interactions in hadron 
collisions since the multiple-radiation effects can be factorized.

Monte Carlo (MC) event generators are indispensable tools for evaluating 
the detection efficiency and acceptance of experiments 
at high-energy hadron colliders.
Since the events in the low-$p_{T}$ peak region dominate the total yield, 
it is important for MC event generators to precisely reproduce 
the spectrum around the peak.
MC event generators simulate the multiple radiation effects with 
parton showers (PS) based on the leading-logarithmic (LL) approximation, 
and non-perturbative effects with certain models.
Thus, their performance is not trivial, particularly around the peak. 
The $Z$-boson $p_{T}$ spectrum is a good test bench for the verification of 
implemented approximations and models.
However, because of the presence of a finite error 
in the momentum measurement of high-momentum decay leptons, 
it is difficult for experiments to make precise measurements 
of the $p_{T}(Z)$ spectrum at low $p_{T}(Z)$.

Alternatively, an angular variable has been introduced for probing 
the transverse activities in $Z$-boson production~\cite{Banfi:2010cf}.
The introduced variable $\phi_{\eta}^{*}$ is defined as 
\begin{equation}\label{eq:phistar}
  \phi_{\eta}^{*} = \tan(\phi_{\rm acop}/2)\sin\theta_{\eta}^{*} . 
\end{equation}
The acoplanarity angle $\phi_{\rm acop}$ is the supplement of the 
opening angle between the two decay leptons in the projection 
onto the plane perpendicular to the beam direction.
The angle $\theta_{\eta}^{*}$ is defined by the pseudorapidity, 
$\eta = -\ln[\tan(\theta/2)]$, of the leptons ($\ell^{\pm}$)
as $\cos\theta_{\eta}^{*} = \tanh[(\eta^{+} - \eta^{-})/2]$.
Although, in general, the momentum resolution deteriorates 
as the $p_{T}$ of charged particles increases, 
the angular resolution usually remains constant irrespective of $p_{T}$.
A transverse recoil of $Z$ bosons more or less produces 
a finite acoplanarity between the decay leptons, 
and the amount of the acoplanarity depends on the decay angle 
in the $Z$-boson rest frame.
The variable $\phi_{\eta}^{*}$ has been proposed to maximize the sensitivity 
to small $Z$-boson recoils,
and it has an approximate correlation to $p_{T}(Z)$ 
as $\phi_{\eta}^{*} \approx p_{T}(Z)/m_{Z}$.

Subsequent to the proposal of this angular variable, 
the D0 experiment at FNAL Tevatron, 
which provides proton-antiproton collisions at a center-of-mass energy 
($\sqrt{s}$) of 1.96 TeV, published their measurement 
on the $\phi_{\eta}^{*}$ spectrum~\cite{Abazov:2010mk}.
They found that the resummation calculation 
by RESBOS~\cite{Balazs:1997xd} provides a good prediction 
with accuracy better than 10\% over almost the entire measurement range.
However, they also found a substantial systematic deviation 
between their measurement and RESBOS.
Although such observations are interesting to be cross-checked, 
D0 has presented the results based on a lepton momentum measurement 
that is not fully corrected for the final-state photon radiation 
(QED FSR) effects in the $Z$ decay.
Hence, an appropriate correction is required in the simulations 
to be compared.
This fact causes a difficulty in the simulation side.

Recently, a new result on the $\phi_{\eta}^{*}$ spectrum 
with significantly higher statistics has been published 
by the ATLAS experiment~\cite{Aad:2012wfa}.
Their measurement is based on proton-proton collision data 
at $\sqrt{s} = 7 {\rm ~TeV}$ provided by CERN LHC.
In contrast to D0, ATLAS has presented the results for several definitions 
of the lepton momenta.
Among them, the Born-level definition, 
in which all QED FSR effects are corrected for in the data analysis, 
allows a straightforward comparison with simulations.
This definition also allows to combine the results for the 
$Z \rightarrow e^{+}e^{-}$ and $Z \rightarrow \mu^{+}\mu^{-}$ channels.
Using the combined high-precision result, 
ATLAS has confirmed the systematic deviation from RESBOS that has been 
observed by D0 with more stringent significance.

In this article, 
we study the performance of the GR@PPA event generator 
by comparing its prediction on the $\phi_{\eta}^{*}$ spectrum 
with the ATLAS measurement.
GR@PPA~\cite{Tsuno:2002ce,Tsuno:2006cu} is an event generator 
for multi-body production processes in hadron collisions 
based on the GRACE system~\cite{Ishikawa:1993qr}.
The version 2.8 package~\cite{Odaka:2011hc,Odaka:2012da} supports 
various weak-boson production processes 
in which associate one-jet production is combined by using a jet-matching 
method~\cite{Kurihara:2002ne,Odaka:2007gu,Odaka:2009qf} 
to reproduce the weak-boson kinematics in the entire phase space.
The package includes custom-made PS programs in order to ensure 
the performance of the jet matching.
We have demonstrated in previous studies~\cite{Odaka:2009qf,Odaka:2012iz} 
that the GR@PPA predictions reproduce the $p_{T}(Z)$ spectra 
measured at Tevatron and LHC with good precision without any tuning.

According to the motivation to introduce the $\phi_{\eta}^{*}$ variable, 
we focus on the spectrum in non-hard regions, $p_{T}(Z) \lesssim m_{Z}$, 
roughly corresponding to $\phi_{\eta}^{*} \lesssim 1.0$. 
The spectrum is predominantly determined by the initial-state PS simulation 
in this region.
The multiple QCD-radiation effects simulated by PS are evaluated by
resummation calculations~\cite{Balazs:1997xd,Bozzi:2010xn} 
on the basis of next-to-leading logarithmic (NLL) or further higher order 
approximations.
Although the theoretical basis of practical PS programs is still limited to 
the LL order, 
higher-order contributions are partially imported in terms of the 
angular ordering in some PS programs~\cite{Corcella:2000bw,Sjostrand:2006za}.
Hence, a question arises: is it necessary to include the effects beyond 
the LL order also in PS simulations for hadron collision interactions?
The answer depends on the accuracy that we require.
Before answering this question, 
we have to investigate how accurately the LL approximation 
can reproduce actual phenomena.
Since the primitive features of the LL approximation are strictly 
preserved in the GR@PPA PS, 
the present study will provide us with an opportunity to probe the capability 
of the LL approximation itself. 

The GR@PPA PS covers the energy region at $Q^{2} > (5 {\rm ~GeV})^{2}$.
Perturbative and non-perturbative contributions at smaller $Q^{2}$ 
are simulated down to the hadron level by feeding the generated 
event information to the general-purpose event generator 
PYTHIA~\cite{Sjostrand:2006za}.
We have found in a previous study~\cite{Odaka:2009qf} 
that the additional PS simulation in PYTHIA is not effective 
in the $p_{T}(Z)$ spectrum, 
whereas the simulation of non-perturbative effects is effective 
around the $p_{T}(Z)$ peak; 
it alters the peak position.
Hence, the verification of the PYTHIA model for non-perturbative effects, 
in the combination with the GR@PPA PS,
is also an important subject in the present study.

The rest of this article is organized as follows. 
The simulation using the GR@PPA event generator is described 
in Sec.~\ref{sec:grappa}.
The result in the default setting is compared with the ATLAS measurement.
Possible alterations in the simulation to achieve better matching 
with the measurement are discussed in Sec.~\ref{sec:optimize}, 
and the discussions are concluded in Sec.~\ref{sec:concl}.

\section{GR@PPA simulation}
\label{sec:grappa}

In the present study, we use the 2.8.4 update of the GR@PPA event generator, 
in which the improvements established 
in the study of diphoton production~\cite{Odaka:2012ry} are migrated 
to the previous 2.8.3 update~\cite{Odaka:2012da}.
However, the impact of this migration is negligible for the $Z$-boson 
production kinematics, 
because the improvements are mainly concerned with the final-state PS.
The event generation is carried out for the 7-TeV LHC condition, 
proton-proton collisions at $\sqrt{s} =$ 7 TeV.
The $Z$ + 0 jet and $Z$ + 1 jet production processes are combined 
by using the jet-matching method.
The decay of the $Z$ boson is included in the matrix elements 
for the event generation, 
and all $Z$ bosons are assumed to decay to an electron pair.
The photon exchange contribution and its interference with the $Z$ exchange 
are also included in the matrix elements. 
The events are generated without any cut, except for the constraint on 
the invariant mass of $Z$ bosons, $66 < m_{Z} < 116 {\rm ~GeV}/c^{2}$.

MRST2007lomod~\cite{Sherstnev:2007nd} 
in the LHAPDF 5.8.4~\cite{Whalley:2005nh} library 
is used for the parton distribution function (PDF) in protons, 
as in the previous study~\cite{Odaka:2012iz}.
The choice of PDF is not important in the simulation of the transverse QCD 
activity, except for the spectrum at very high $p_{T}$ ($\gtrsim m_{Z}$) 
where the $Z$ + 1 jet contribution becomes dominant.
The initial-state QCD radiations are simulated by using the backward-evolution 
PS (QCDPSb) included in the GR@PPA package.
The energy scales, the factorization scale ($\mu_{F}$) and the renormalization 
scale ($\mu_{R}$), are chosen as $\mu_{F} = \mu_{R} = m_{Z}$.
The energy scale of the initial-state PS is always equal to $\mu_{F}$ 
in our jet-matching method.
The final-state PS (QCDPSf) is also activated.
Although it is not important in the present study, 
the final-state PS energy scale is also set to $m_{Z}$.

The generated events are exported to PYTHIA 6.425~\cite{Sjostrand:2006za} 
in order to add simulations at small $Q^{2}$ along with simulations 
of hadronization and decays.
The default setting in PYTHIA is unchanged except for the setting of 
{\tt PARP(67) = 1.0} and {\tt PARP(71) = 1.0}, 
as in the previous studies~\cite{Odaka:2009qf,Odaka:2012iz}.
The ATLAS measurement is presented for the quantities of Born-level leptons, 
in which the QED FSR effect in the $Z$ decay is corrected for.
In order to follow this definition, 
we look for an electron pair from a $Z$-boson decay 
before the QED FSR simulation in the PYTHIA event record.
The pair is the same one as that generated by GR@PPA, 
but their rest frame is boosted and rotated by the simulations in PYTHIA.
Since these electrons are defined at the Born level, 
they can be replaced with muons.
Hence, we refer to these electrons as generic leptons ($\ell^{\pm}$) 
in the following discussions.

The event selection is applied using the momenta of these leptons 
according to the definition of ATLAS; 
that is, $p_{T} > 20 {\rm ~GeV}/c$ and $|\eta| < 2.4$ are required 
for both leptons.
The invariant mass is already constrained in the event generation.
The selected events are binned according to the binning 
on $\phi_{\eta}^{*}$ defined by ATLAS
to derive the $(1/\sigma)d\sigma/d\phi_{\eta}^{*}$ spectrum.

\begin{figure}[tp]
\begin{center}
\includegraphics[scale=0.6]{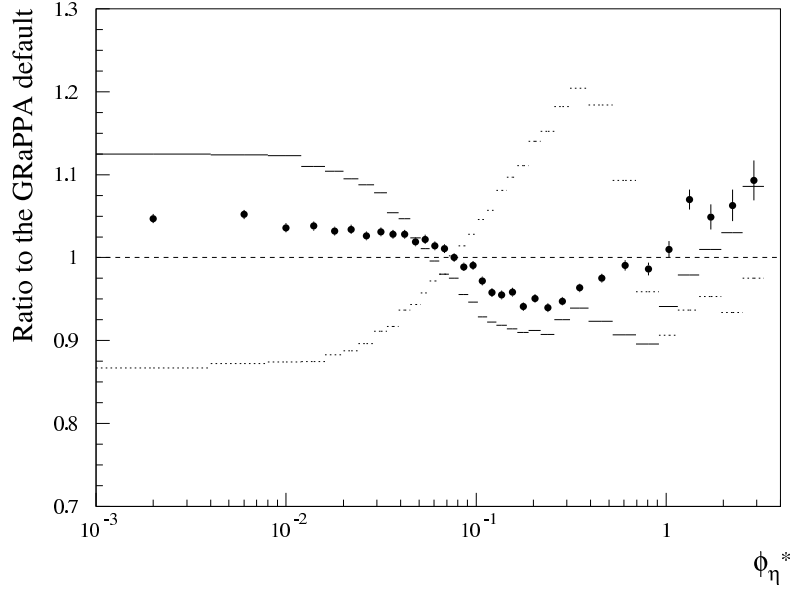}
\caption{\label{fig:default}
$(1/\sigma)d\sigma/d\phi_{\eta}^{*}$ spectrum normalized to 
the prediction from the GR@PPA simulation with the default setting.
The result for the spectrum measured by ATLAS is plotted.
The solid and dotted bars illustrate the predictions from GR@PPA 
with extreme settings of the arbitrary energy scales: 
solid bars for $\mu_{F} = \mu_{R} = 0.5m_{Z}$, 
and dotted bars for $\mu_{F} = \mu_{R} = 2.0m_{Z}$.
}
\end{center}
\end{figure}

Figure~\ref{fig:default} shows the ratio of the ATLAS measurement 
with respect to the GR@PPA simulation result.
The results for the $Z \rightarrow e^{+}e^{-}$ and 
$Z \rightarrow \mu^{+}\mu^{-}$ channels are combined 
in the ATLAS measurement.
We can see the agreement fairly better than 10\% between the simulation 
and measurement over the entire measurement range.
However, we also see an apparent systematic deviation between them.
The tendency is similar to that observed in the comparison 
with RESBOS in the ATLAS paper~\cite{Aad:2012wfa}, 
although the deviation is slightly larger in the present result.

The most remarkable feature of the result in Fig.~\ref{fig:default} 
is that the measurement/simulation ratio is significantly smaller than unity 
at medium $\phi_{\eta}^{*}$, $0.1 \lesssim \phi_{\eta}^{*} \lesssim 0.5$.
This means that the event fraction in the medium $p_{T}(Z)$ range, 
$10 \lesssim p_{T}(Z) \lesssim 50 {\rm ~GeV}/c$, 
is too large in the simulation. 
The fraction becomes smaller than the measurement in the peak region, 
$\phi_{\eta}^{*} \lesssim 0.1$, to compensate for the enhancement 
since the compared distributions are normalized.
We found in a previous study~\cite{Odaka:2009qf} 
that the $p_{T}(Z)$ spectrum at $p_{T}(Z) \gtrsim 10 {\rm ~GeV}/c$ 
is not altered by the PYTHIA simulation. 
Hence, the observation implies that the PS radiation effect is 
too strong in the GR@PPA event generation.

\begin{figure}[tp]
\begin{center}
\includegraphics[scale=0.6]{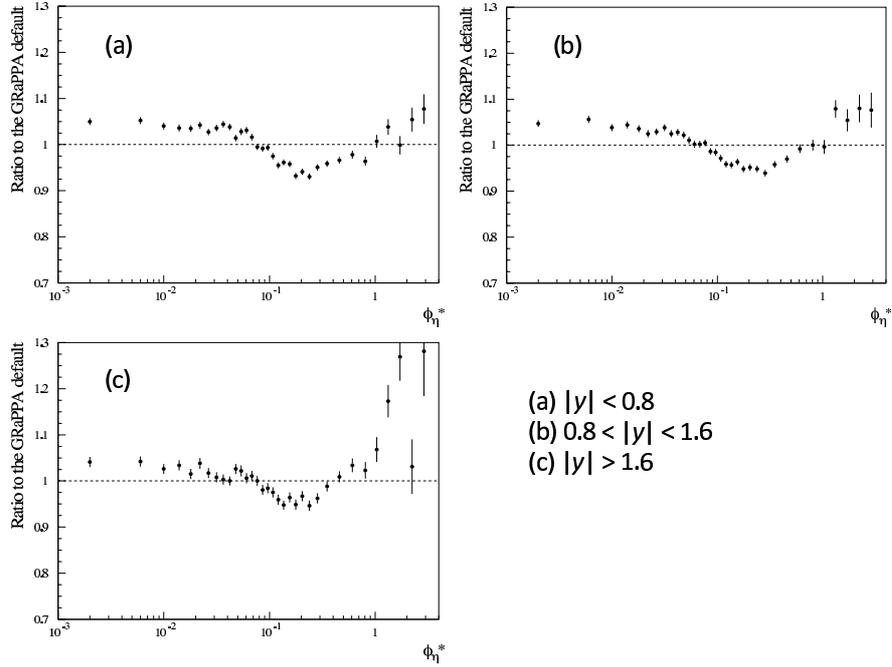}
\caption{\label{fig:default_y}
$(1/\sigma)d\sigma/d\phi_{\eta}^{*}$ spectrum normalized to 
the prediction from the GR@PPA simulation with the default setting.
The result for the ATLAS measurement is separately plotted 
in three rapidity regions of the lepton pair: 
(a) $|y_{\ell\ell}| < 0.8$, (b) $0.8 < |y_{\ell\ell}| < 1.6$, 
and (c) $|y_{\ell\ell}| > 1.6$.
}
\end{center}
\end{figure}

Figure~\ref{fig:default_y} shows the result of the comparison in 
three rapidity regions of the lepton pair.
The rapidity is defined by the total energy $E$ and the momentum 
along the beam direction $p_{z}$ as $y = (1/2)\ln[(E+p_{z})/(E-p_{z})]$.
We can see that the overall behavior is almost independent of the rapidity, 
except in the large $\phi_{\eta}^{*}$ region, 
$\phi_{\eta}^{*} \gtrsim 1.0$, 
where the $Z$ + 1 jet contribution is dominant.
The gluon density in PDF contributes to the $Z$ + 1 jet cross section 
together with quark densities, 
whereas the total yield is predominantly determined by quark densities.
Thus, the fraction in the large $\phi_{\eta}^{*}$ region should be 
altered by the change in the gluon/quark ratio in PDF 
in the relevant parton momentum range.
The observation at large $\phi_{\eta}^{*}$ would provide additional 
information on PDF.
However, we do not discuss further details of this issue 
in this article.
Detailed discussions should be carried out using the $p_{T}(Z)$ 
distribution.

It is to be noted that the error bars illustrated 
in Figs.~\ref{fig:default} and \ref{fig:default_y} 
show the quadratic sum of all errors: 
the statistical and systematic errors of the measurement 
and the statistical error of the simulation.
The simulation is based on $10^{7}$ events generated by GR@PPA, 
and the number of events is reduced to about $5 \times 10^{6}$ 
after the event selection.
Because the ATLAS measurement is based on a huge amount of data, 
approximately $3 \times 10^{6}$ lepton-pair events, 
the contribution of the simulation statistics is not negligible 
in the illustrated errors.
However, this contribution does not affect the above discussions 
since the observed systematic deviation is considerably larger 
than the estimated error.

\section{Optimization of the simulation}
\label{sec:optimize}

The $Z$ production cross section evaluated on the basis of 
the lowest-order $q\bar{q} \rightarrow Z$ interaction 
necessarily has a dependence on the factorization scale ($\mu_{F}$),
since PDFs are provided as a function of $\mu_{F}$. 
In principle, this scale can be chosen arbitrarily.
This dependence is mostly canceled by the inclusion of $Z$ + 1 jet 
production in our jet-matching method~\cite{Kurihara:2002ne,Odaka:2011hc}, 
and the inclusion of the $Z$ + 1 jet extends the $p_{T}(Z)$ spectrum 
to high $p_{T}(Z)$ to cover the entire phase space.
We can achieve good matching between the $Z$ + 0 jet and $Z$ + 1 jet 
in the $p_{T}(Z)$ spectrum by introducing an appropriate parton branch 
model~\cite{Odaka:2007gu,Odaka:2009qf}.
However, because the cancellation is realized only 
at the first order of the QCD coupling, $\alpha_{s}$, 
a substantial dependence remains not only in the total cross section 
but also in the $p_{T}(Z)$ spectrum~\cite{Odaka:2007gu,Odaka:2009qf}.

The evaluated cross section also has an ambiguity due to the arbitrariness 
in the renormalization scale ($\mu_{R}$). 
The $\alpha_{s}$ value in matrix-element (ME) calculations 
depends on $\mu_{R}$. 
Since $q\bar{q} \rightarrow Z$ is an electroweak interaction, 
the ambiguity affects the $Z$ production cross section 
only through the $Z$ + 1 jet production.
Thus, the effect is small but substantial.

These ambiguities depending on the choice of the energy scales, 
$\mu_{F}$ and $\mu_{R}$, are caused by the lack of higher-order contributions.
The variation in the simulation result reflects 
the uncertainty in the theory that the simulation is based on.
In our jet-matching method, we take $\mu_{R} = \mu_{F}$ 
in order to equalize the radiation probability in the PS applied 
to the $Z$ + 0 jet and in the MEs for the $Z$ + 1 jet at their boundary, 
$Q^{2} = \mu_{F}^{2}$.
The energy scales are usually set to the "typical" energy 
of the interaction of interest, 
because non-collinear contributions ignored in PS and PDF become 
significant and higher-order contributions are minimized around this energy. 
Hence, we take $\mu_{R} = \mu_{F} = m_{Z}$ as the default for $Z$ production, 
as described in the previous section.

Theoretical uncertainties are usually evaluated by multiplying 
the default choice of the energy scales by factors of 0.5 and 2.0.
The results for these extreme choices, $\mu_{R} = \mu_{F} = 0.5m_{Z}$ 
and $\mu_{R} = \mu_{F} = 2.0m_{Z}$, are illustrated in Fig.~\ref{fig:default} 
using solid and dashed bars, respectively.
We can see that the ATLAS measurement lies within the range of 
the uncertainty represented by these two results in the range 
$\phi_{\eta}^{*} \lesssim 1.0$.

The theoretical uncertainty is rather large because the MEs are evaluated 
only at the tree level.
However, the main purpose of MC event generators is not to provide 
theoretical predictions, but to provide a tool for analyses in experiments.
The simulations are desired to reproduce actual measurements 
as precisely as possible in this usage.
Adjustments of parameters in the simulation should be allowed to satisfy 
this requirement.
Of course, the adjustments have to be carried out in a reasonable range 
or within the range that the underlying theory allows.
Looking at the results in Fig.~\ref{fig:default}, 
we speculate that better matching may be achieved 
with energy scales slightly smaller than the default values, 
although it is not trivial that the observed systematic deviation vanishes 
with such a simple adjustment.

\begin{figure}[tp]
\begin{center}
\includegraphics[scale=0.6]{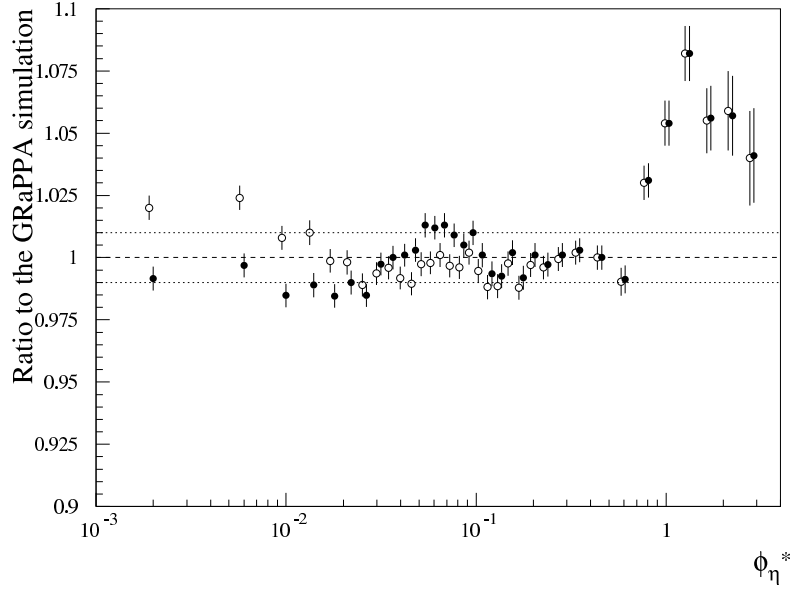}
\caption{\label{fig:adjust}
$(1/\sigma)d\sigma/d\phi_{\eta}^{*}$ spectrum normalized to 
the prediction from the GR@PPA simulation with the adjusted energy scales, 
$\mu_{R} = \mu_{F} = 0.75m_{Z}$.
Two results for the ATLAS measurement are plotted. 
Filled circles show the ratio to the simulation employing 
the default primordial $k_{T}$ simulation, $k_{T} = 2.0 {\rm ~GeV}$, 
in PYTHIA,
while open circles show the ratio to the simulation 
with $k_{T} = 2.4 {\rm ~GeV}$.
Dotted lines indicate the $\pm$1\% deviation between the measurement 
and simulation.
}
\end{center}
\end{figure}

\begin{figure}[tp]
\begin{center}
\includegraphics[scale=0.6]{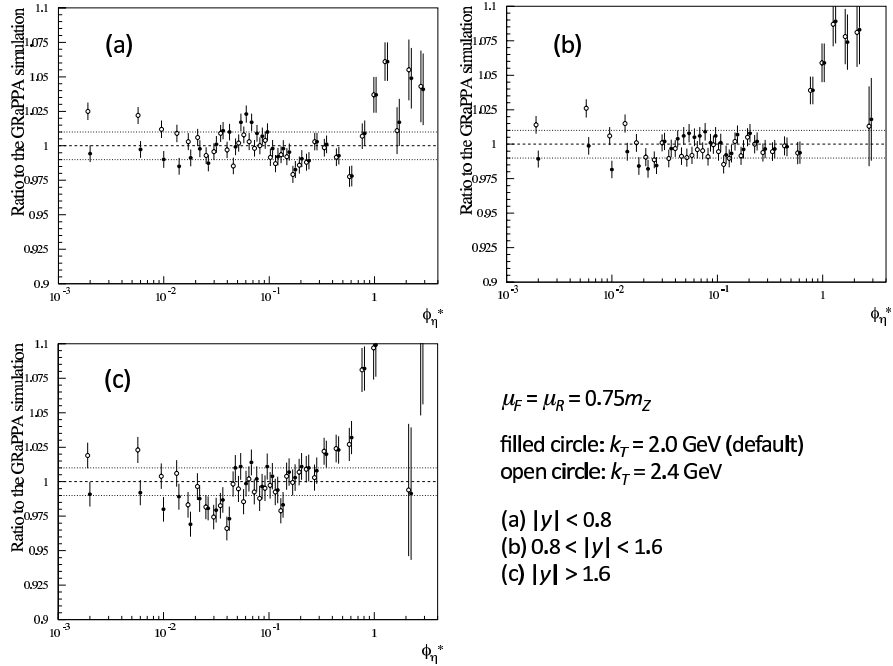}
\caption{\label{fig:adjust_y}
$(1/\sigma)d\sigma/d\phi_{\eta}^{*}$ spectrum normalized to 
the prediction from the GR@PPA simulation with the adjusted energy scales, 
$\mu_{R} = \mu_{F} = 0.75m_{Z}$.
The notation is the same as that in Fig.~\ref{fig:adjust}, 
but the results are separately shown in three rapidity regions 
of the lepton pair: 
(a) $|y_{\ell\ell}| < 0.8$, (b) $0.8 < |y_{\ell\ell}| < 1.6$, 
and (c) $|y_{\ell\ell}| > 1.6$.
Dotted lines indicate the $\pm$1\% deviation between the measurement 
and simulation.
}
\end{center}
\end{figure}

We have tried several settings and found that 
we can obtain an almost optimum result with the choice 
of $\mu_{R} = \mu_{F} = 0.75m_{Z}$.
The result with this setting is plotted in Fig.~\ref{fig:adjust} 
with filled circles.
In this result, the simulation statistics are improved by 
a factor of five with respect to the study in the previous section.
The resultant statistical error is approximately 1/3 of that 
from the measurement; thus, it is almost negligible.
We can see that the dip that is observed in Fig.~\ref{fig:default} 
in the medium-$\phi_{\eta}^{*}$ region, $0.1 < \phi_{\eta}^{*} < 0.5$, 
has vanished. 
The simulation matches the measurement with accuracy 
better than 2\% through the measurement range 
up to $\phi_{\eta}^{*} \approx 1.0$.
Note that the vertical scale is expanded in this plot 
with respect to Fig.~\ref{fig:default}.
Figure~\ref{fig:adjust_y} shows the results 
in the three rapidity ($y_{\ell\ell}$) ranges.
We can see that the adjustment has reasonably improved the matching 
in all $y_{\ell\ell}$ bins.

In spite of the observed overall improvement, 
the measurement/simulation ratio still shows a substantial systematic 
discrepancy from unity in the small-$\phi_{\eta}^{*}$ region, 
$\phi_{\eta}^{*} < 0.1$.
This small-$\phi_{\eta}^{*}$ region corresponds to the peak region 
in the $p_{T}(Z)$ spectrum, $p_{T}(Z) \lesssim 10 {\rm ~GeV}$, 
where simulations for non-perturbative effects in PYTHIA 
based on certain models are effective.
Among various simulations applied in PYTHIA, 
the primordial-$k_{T}$ simulation plays the most 
important role for determining the $p_{T}(Z)$ spectrum 
in this region~\cite{Odaka:2009qf}.
This simulation is considered to include non-perturbative effects 
together with small-$Q^{2}$ perturbative effects below the PYTHIA-PS 
cutoff, $Q_{0}^{2} = (1.0 {\rm ~GeV})^{2}$.
The applied model is very simple; 
an additional $p_{T}$ is assumed for each parton in protons according to 
the Gaussian distribution having the standard deviation of $k_{T}$.
The default setting is $k_{T} = 2.0 {\rm ~GeV}$.

The result in Fig.~\ref{fig:adjust} shown with filled circles indicates 
that the PYTHIA model for the primordial $k_{T}$ is almost appropriate. 
However, there is still some room for optimizing the $k_{T}$ parameter.
The open circles in Fig.~\ref{fig:adjust} show the result 
when we increase the $k_{T}$ value to 2.4 GeV.
The plots are slightly shifted to the left 
in order to ensure that the changes are visible.
Although the matching deteriorates 
at very small $\phi_{\eta}^{*}$ ($< 0.01$), 
where even the angular measurement may be ambiguous, 
a substantial improvement can be seen in the region of 
$0.01 < \phi_{\eta}^{*} < 0.1$.
As a result, the measurement/simulation ratio distribution becomes 
flat within $\pm 1\%$ over a very wide range, 
$0.01 \lesssim \phi_{\eta}^{*} \lesssim 1.0$.
The results in three separate $y_{\ell\ell}$ regions 
after the $k_{T}$ adjustment are also illustrated with open circles 
in Fig.~\ref{fig:adjust_y}.
We can see substantial improvements that are similar to the improvement 
observed in Fig.~\ref{fig:adjust}.

\section{Conclusion}
\label{sec:concl}

The measurement of the angular variable $\phi_{\eta}^{*}$, 
which is defined for the production of $Z$ bosons in hadron collisions 
decaying to a lepton pair, provides us with rich information 
concerning the initial-state QCD activities, 
especially at small $p_{T}(Z)$ where the $p_{T}(Z)$ measurement 
is ambiguous.
It has been demonstrated in this article that the normalized 
$\phi_{\eta}^{*}$ spectrum, $(1/\sigma)d\sigma/d\phi_{\eta}^{*}$, 
measured by the ATLAS experiment at LHC can be reproduced 
by the simulation based on the GR@PPA event generator 
with the precision better than 5\% over almost the entire measurement 
range where the spectrum is predominantly determined by multiple 
QCD radiations simulated by PS.

Simulations based on MC event generators are desired to reproduce 
measurement data as precisely as possible 
when they are used as tools for measurements.
We have shown that the observed disagreement between the measurement 
and simulation is well within the theoretical uncertainty, 
which is represented by the variation due to the change 
of arbitrary energy scales in the event generation, 
and we have demonstrated that the agreement can be improved 
to the level of 1\% by adjusting these scales.
It should be emphasized that this level of precision has been 
achieved with a primitive leading-logarithmic PS 
implemented in the GR@PPA event generator.

We use the PYTHIA event generator for simulating non-perturbative 
and perturbative effects at small $Q^{2}$ ($< (5 {\rm ~GeV})^{2}$) 
in our simulation.
The successful simulation at small $\phi_{\eta}^{*}$ 
implies that the simulation in PYTHIA is basically appropriate.
However, the remaining fine structure in the measurement/simulation 
ratio suggests the necessity of a fine tuning of the $k_{T}$ parameter 
in the primordial-$k_{T}$ simulation in PYTHIA.
The change of the parameter $k_{T}$ to 2.4 GeV from its default value 
of 2.0 GeV improves the matching in the relevant $\phi_{\eta}^{*}$ range.

In this article, 
we have discussed the capability of the GR@PPA event generator 
as a tool for $Z$-boson production measurements.
We anticipate that a similar level of precision can be achieved for 
other processes, 
although the optimum choice of the parameters would be dependent 
on the process because the higher-order contributions would be different.
The collision-energy dependence of the $k_{T}$ parameter may also 
be an interesting subject for future investigation.
The GR@PPA event generator is being developed to provide a tool 
for such applications and studies.
The most up-to-date version of the package is available from the Web 
page\footnote{\tt http://atlas.kek.jp/physics/nlo-wg/grappa.html.}.

\section*{Acknowledgments}

This work has been carried out as an activity of the NLO Working Group, 
a collaboration between the Japanese ATLAS group and the numerical analysis 
group (Minami-Tateya group) at KEK.
The author wishes to acknowledge useful discussions with the members, 
especially Y. Kurihara and N. Watanabe.

\providecommand{\href}[2]{#2}\begingroup\raggedright\endgroup

\end{document}